\documentclass[showpacs,twocolumn,pre]{revtex4}

\usepackage[dvips]{epsfig}

\newcommand{\clG}{{\cal G}}
\newcommand{\clP}{{\cal P}}
\newcommand{\clF}{{\cal F}}

\newcommand{\clL}{{\cal L}}

\newcommand{\hL}{\hat{L}}
\newcommand{\hH}{\hat{\cal H}_0}
\newcommand{\PBr}{\{H,\dots\}_{PB}}
\newcommand{\clE}{{\cal E}}

\newcommand{\prt}{\partial}

\newcommand{\rgl}{\rangle}
\newcommand{\lgl}{\langle}

\newcommand{\be}{\begin{equation}}
\newcommand{\ee}{\end{equation}}
\newcommand{\bea}{\begin{eqnarray}}
\newcommand{\eea}{\end{eqnarray}}

\begin{document}

\title{Subdiffusion in the Nonlinear
Schr\"odinger Equation with Disorder}

\author{Alexander Iomin}

\affiliation{Department of Physics, Technion, Haifa, 32000,
Israel}
\date{\today}

\begin{abstract}
The nonlinear Schr\"odinger equation (NLSE) in the presence of
disorder is considered. The dynamics of an initially localized
wave packet is studied. A subdiffusive spreading of the wave
packet is explained in the framework of a continuous time random
walk. A probabilistic description of subdiffusion is  suggested
and a transport exponent of subdiffusion is obtained to be $2/5$.

\end{abstract}

\pacs{05.45.Yv, 72.15.Rn, 42.25.Dd}

\maketitle

In this work the dynamics of an initially localized wave function
is considered in the framework of the nonlinear Schr\"odinger
equation in the presence of disorder. The wave function is
governed by the following equation
\be\label{am1}
 i\prt_t\psi=-\prt_x^2\psi+\beta|\psi|^2\psi+V\psi\, ,
\ee %
where $\beta$ is a nonlinearity parameter.  The variables are
chosen in dimensionless units and the Planck constant is
$\hbar=1$. The random potential $V=V(x), ~x\in(-\infty,+\infty)$
is such that for the linear case $(\beta=0)$ the Anderson
localization takes place \cite{Anderson,Lee}, and the system is
described by the exponentially localized Anderson modes (AM)s
$\Psi_{\omega_k}\equiv\Psi_k(x)$, where $\Psi_{\omega_k}(x)$ are
real functions and the eigenspectrum $\omega_k$ is discrete and
dense \cite{LGP}. The problem in question is an evolution of an
initially localized wave function $\psi(t=0)=\psi_0(x)$.

This problem is relevant to experiments in nonlinear optics, for
example disordered photonic lattices \cite{f1,lahini}, where
Anderson localization was found in the presence of nonlinear
effects, as well as to experiments on Bose-Einstein Condensates in
disordered optical lattices
\cite{BECE1,BECE3,akkermans,SanchAspect,aspect,BShapiro}. A
discrete analog of Eq. (\ref{am1}) is extensively studied
numerically \cite{dima,molina,Pikovsky,fks}, and a subdiffusive
spreading of the initially localized wave packet has been
observed, such that $\lgl x^2(t)\rgl=\int|\psi(t)|^2x^2dx\sim
t^{\alpha}$, where a transport exponent $\alpha$ was found in the
range $0.3\div 0.4$ \cite{Pikovsky,fks}.

A subdiffusive spreading of the wave packet was also obtained
analytically \cite{iom1} in the limit of the large times
asymptotic dynamics of the tails of the wave packet, where the
transport exponent was found to be $\alpha<1$. In that case the
dynamics of the wave packet has been approximated by the
fractional Fokker-Planck equation (FFPE) due to the arguments of a
so-called continuous time random walk (CTRW).

The concept of the CTRW was originally developed for mean first
passage time in a random walk on a lattice with further
application to a semiconductor electronic motion \cite{montweiss}.
The mathematical apparatus of the fractional CTRW is well
established for many applications in physics, see {\it e.g.},
\cite{bouchaud,klafter,zaslavsky,benAvrHavlin,shlesinger}.

The primary purpose of the present analysis is to develop  the
physical mechanism of the wave packet spreading and to obtain the
transport exponent $\alpha$. The analysis is based on
mapping the nonlinear Eq. (\ref{am1}) to the linear Liouville
equation for the probability amplitude $|\psi(x,t)|^2$, where the
transition elements in the Liouville operator are determined by
the nonlinear term in Eq. (\ref{am1}). Therefore we proceed to
developing the analysis further by application of the CTRW
approach to the corresponding Liouville equation \cite{kenkre}.

First we obtain the linear Liouville equation for $|\psi(x,t)|^2$ \cite{iom1}.
Projecting Eq. (\ref{am1}) on the basis of the AMs
\be\label{am2}
 \psi(x,t)=\sum_{\omega_k}C_{\omega_k}(t)
\Psi_{\omega_k}(x)\equiv\sum_kC_k(t)\Psi_k(x)\, ,
 \ee %
we obtain a system of equations for coefficients of the expansion
$C_k$
\be\label{am3}
 i\prt_t{C}_k=\omega_kC_k+
\beta\sum_{k_1,k_2,k_3}A_{k_2,k_3}^{k,k_1}
C_{k_1}^*C_{k_2}C_{k_3}\, .
 \ee %
Here $A({\bf k})\equiv A_{k_2,k_3}^{k,k_1}$  is an overlapping
integral of the four AMs:
\be\label{am4}
 A_{k_2,k_3}^{k,k_1}
=\int\Psi_k(x)\Psi_{k_1}(x)\Psi_{k_2}(x)\Psi_{k_3}(x)dx\, .
 \ee%
The initial conditions for the system of Eqs. (\ref{am3}) are such
that $ \psi_0(x)=\sum_ka_k\Psi_k(x)$. Equations (\ref{am3})
correspond to a system of interacting nonlinear oscillators with
the Hamiltonian
\be\label{am6}
 H=\sum_k\omega_kC_k^*C_k+\beta\sum_{\bf
k}A_{k_2,k_3}^{k_1,k_4}
 C_{k_1}^*C_{k_4}^*C_{k_2}C_{k_3}\, .
\ee %
Therefore, Eqs. (\ref{am3}) are produced by the Poisson brackets
$\PBr$ by means of the Liouville operator
\be\label{am7}
 \hL=\frac{1}{i}\PBr  
= \frac{1}{i}\sum_k\left(\frac{\prt H}{\prt
C_k^*}\cdot\frac{\prt}{\prt C_k}-\frac{\prt H}{\prt
C_k}\cdot\frac{\prt}{\prt C_k^*}\right)\, .
 \ee %
Since $\hL H=0$ and $H(\{C,C^*\})=H(\{a,a^*\})$, we obtain that
the Liouville operator is an operator function of the initial
values:
\be\label{am7a}%
\hL=\frac{1}{i}\sum_k\left[\frac{\prt H}{\prt
a_k^*}\cdot\frac{\prt}{\prt a_k} -\frac{\prt H}{\prt
a_k}\cdot\frac{\prt}{\prt a_k^*}\right]
\ee%
and corresponds to an infinite system of linear equations $\prt_t{\bf
C}=\hL{\bf C}$, where ${\bf C}={\bf C}(\{a_k,a_k^*\})=(\dots ,
C_{k-1},C_k,C_{k+1},\dots)$ is an infinite vector. Thus, the
Liouville operator reads
\bea\label{am7b}%
\hL=&-&i\sum_k\omega_k\left(a_k\frac{\prt}{\prt a_k}-{\rm
c.c.}\right) \nonumber  \\
&-&i\beta\sum_{\bf k} A_{k_2,k_3}^{k_1,k_4}
\left[a_{k_1}^*a_{k_2}a_{k_3}\frac{\prt}{\prt a_{k_4}} - {\rm
c.c.}\right]\, ,   %
\eea%
where c.c. denotes the complex conjugation. The Liouville equation
is valid for any functions of the initial conditions
$\{a_k,~a_k^*\}$. In particular, introducing the function
$F_{k,k'}(t)=C_k(t)\cdot C_{k'}^*(t)$, one has the Liouville
equation:
\[\prt_tF_{k,k'}(t)=\hL F_{k,k'}(t)\, ,
~~~F_{k,k'}(t=0)=F_{k,k'}^{(0)}=a_ka_{k'}^*\, .\]
Therefore, the probability amplitude
\[\clP(x,t)=|\psi|^2=\sum_{k,k'}F_{k,k'}(t)\Psi_k(x)\Psi_{k'}(x)\, ,\]
as a function of the initial conditions,  satisfies the Liouville
equation as well:
\be\label{trueWF}   %
\prt_t\clP=\hL\clP\, .
\ee %
Here the initial condition is
\be\label{icond} %
\clP(x,t=0)=\clP_0(x)=\sum_{k,k'}F_{k,k'}^{(0)}\Psi_k(x)\Psi_{k'}(x)\, .  %
\ee  %
Finally, we obtain that the NLSE (\ref{am1}) is replaced by the
Liouville equation (\ref{trueWF}), which is the \textit{linear}
equation with a formal solution in the exponential form
\be\label{am10_a} %
\clP(x,t)=e^{\hL t}\clP_0(x)=
\sum_{k,k'}\Psi_k(x)\Psi_{k'}(x)\sum_{n=0}^{\infty}\Big[\frac{t^n}{n!}
\hL ^n\Big]F_{k,k'}^{(0)}\, .  %
\ee %

In what follows we consider the dynamics of the initial wave
packet $\clP_0(x)$ in the framework of the probabilistic approach,
where the dynamics of the wave packet is considered as the CTRW.
Since the dynamics of the probability distribution function (PDF)
$\clP(x,t)$ is governed by the same
Liouville operator in Eqs. (\ref{trueWF}) and (\ref{am10_a}), the
overlapping integrals $A({\bf k})$ play the dominant role in the
wave packet spreading. As follows from Eq. (\ref{am10_a}), the
overlapping integrals determine the spread of the initially
localized wave packet $\clP_0(x)$ over all the Anderson modes as
transitions from one mode to another. Since all states are
localized, these transitions between states determine the
transitions (or jumps) in the coordinate space as well.

These transitions due to the
overlapping integrals can be considered in the framework of the formal
probabilistic integral equation (\ref{trueWF}). Than the  formal
solution of Eq. (\ref{am10_a}) can have a form of the master
equation
\be\label{masteq}  %
\clP(x,t)=\int_0^t\int_{-\infty}^{\infty}\clG(x,t;x't')\clP(x',t')dx'dt'\,
,\ee %
where the Green function $\clG(x,t;x't')$ can be determined
from the analysis of the overlapping integrals based on the
probabilistic approach. \cite{add1}.

According to the values of the overlapping integrals, we divide
the transitions between the localized states into two main groups.
The first one corresponds to the exponentially small overlapping
integrals and the second one corresponds to the strong overlapping
between four AMs when the overlapping integrals are of the order
of $1$.  In the case of strong overlapping, the AMs form clusters,
where the wave functions have the same averaged coordinates for
each cluster. Consequently, all transitions inside one cluster do
not lead to any appreciable differences in the coordinate space,
and we regard these transitions as \textit{trapping} of the
wave packet, or a particle, inside this cluster. Contrary to that,
transitions due to the exponentially small overlapping integrals
between the AMs belonging to different clusters lead to a change
of the space coordinates that can be accounted for. We call these
processes \textit{jumps}. Contributions of trappings and jumps
to the wave packet spreading described by Eqs. (\ref{trueWF}) and
(\ref{am10_a}) are different, and correspond to different
probabilistic interpretations.

In the sequel we follow the CTRW approach \cite{montweiss}
and paraphrase it from \cite{klafter,benAvrHavlin} in a form suitable for
the present analysis. First, we consider a process of jumps.
Let $P_n(x)$ be the PDF of being at
$x$ after $n$ jumps. It is reasonable to assume that the transitions
between different states are independent of each other; therefore, the jumps are
independent and obey the Markov property
\be\label{markov_j} %
P_{n+1}(x)=\int P_n(x')p(x-x')dx'\, ,   %
\ee %
where $P_0(x)=\clP_0(x)$ and $p(x)$ is the PDF of a jump
determined by the overlapping integrals as
$p(x)=\xi\exp(-\xi|x|)/2$, and $\xi=1/R$ is an inverse
localization length.

The trapping is associated with clusters with effective lengths
$\Delta$. Due to the exponential localization, these values are
distributed by the exponential law $P_{\rm cl}\left(\Delta
\right)=\Delta_0^{-1}\exp\left(-\Delta /\Delta_0\right)$, where
$\Delta_0$ is the mean length of a cluster. The effective lengths
are determined by overlapping integrals; therefore, the minimum
length of the cluster is $\Delta =R$, while the maximum one is
$\Delta =4R$. As a result, one obtains $\Delta_0=5R/2$. Now the
probability that a particle exits this cluster and jumps to
another one is of the order of $\sim \exp\left(-\Delta /R\right)$.
This value is also proportional to the inverse waiting time,
$t\sim\exp\left(\Delta /R\right)$. The probability to find the
waiting time in the interval $(t,~t+dt)$ is equal to the
probability to find the corresponding trapping length in the
interval $(\Delta,~\Delta+d\Delta)$. Therefore the PDF of the
waiting times is
\be\label{pdf_a} %
w(t)=P_{\rm cl}\left(\Delta \right)\frac{d \Delta }{dt} \sim
\frac{1}{(t/\tau)^{1+\alpha}}\, ,       %
\ee    %
where $\alpha=2/5$ and $\tau$ is a time scale related to the
trapping \cite{add3}. It follows that the mean waiting time is infinite.
Taking into account that the waiting time PDF is normalized we
have
\be\label{pdf_w} %
w(t)=\frac{w_0}{1+(t/\tau)^{1+\alpha}}\, ,    %
\ee %
such that $ \int_0^{\infty}w(t)dt =1$ and
$\int_0^{\infty}tw(t)dt=\infty$. Here
$w_0=\frac{2\alpha\sin(\pi/2\alpha)}{\tau^{(1+\alpha)/2\alpha}}$
is a normalization constant.

Now let us consider the PDF $w(t)$ taking into account the
dynamics of the jumps. Again, since transitions between different
states are statistically independent, waiting times for different
jumps are statistically independent as well. Therefore, indexing
the waiting time PDF by the jump number, we define that $w_n(t)$
is the probability density that $n$th jump occurs at time $t$ (see
\textit{e.g.}, \cite{benAvrHavlin}, p.42).  Due to the reasonable assumption
that jumps are independent transitions, we also introduce the Markov property
for $w_n(t)$, which reads
\be\label{markov_w} %
w_{n+1}(t)=\int_0^{\infty}w_n(t')w(t-t')dt'\, ,   %
\ee  %
where $w_1(t)\equiv w(t)$. Now we introduce the PDF
$P(x,t)=\sum_nP_n(x)w_n(t)$ of arriving at coordinate $x$ at time
$t$. From Eqs. (\ref{markov_j}) and (\ref{markov_w}) we introduce
the equation \cite{klafter}
\bea\label{inteq_a}  %
P(x,t)&=&\int_{-\infty}^{\infty}p(x-x')\int_0^{\infty}w(t-t')
P(X',t')dx'dt' \nonumber  \\
&+&\clP_0(x)\delta(t)\,
, \eea   %
which relates the PDF $P(x,t)$ of just having arrived at position
$x$ at time $t$ to the PDF $P(x',t')$ of just arriving at $x'$
at time $t'$. The last term in Eq. (\ref{inteq_a}) is the initial
condition. Thus the PDF $\clP(x,t)$ of being at position $x$ at
time $t$ is given by arrival at $x'$ at time $t'$ and not having
moved after this event, namely
\be\label{pdf_clP}   %
\clP(x,t)=\int_0^tP(x,t')W(t-t')dt'\, ,   %
\ee %
where $W(t)=1-\int_0^tw(t')dt'$ denotes the probability of no jump
during the time interval $(0,t)$. Performing the Fourier transform
$\bar{p}(k)=\hat{\clF}p(x)$ and the Laplace transform
$\tilde{w}(s)=\hat{\clL}w(t)$, we obtain the Montroll-Weiss equation
\cite{montweiss}
\be\label{MW}  %
\bar{\tilde{\clP}}(k,s)=\hat{\clF}\hat{\clL}\clP=
\frac{1-\tilde{w}(s)}{s}\cdot\frac{\bar{\clP}_0(k)}{1-\bar{p}(k)\tilde{w}(s)}\,
. \ee  %
This expression determines the master equation (\ref{masteq}) and
establishes a relation between the Green function and the overlapping
integrals in the CTRW form: $\clG=p(x-x')w(t-t')$. Eq. (\ref{MW}) can be
simplified for the long time
$s\ll 1$ and the large scale $k\ll 1$ asymtotics that corresponds to the
diffusion limit $(k,s)\rightarrow(0,0)$. Taking into account the
Fourier $\bar{p}(k)$ and the Laplace $\tilde{w}(s)$ images in Eq.
(\ref{MW}):
\bea\label{asympt}  %
\bar{p}(k)&=&\frac{1}{1+R^2k^2}\approx 1-R^2k^2\,  , \nonumber \\
\tilde{w}(s)&=&\frac{1}{1+(s\tau)^{\alpha}}\approx  1-(s\tau)^{\alpha}\, ,
\eea  %
we obtain for the
PDF in the Fourier-Laplace domain (see also \cite{klafter})
\be\label{MWasy}      %
\bar{\tilde{\clP}}(k,s)=
\frac{\bar{\clP}_0(k)/s}{1+D_{\alpha}s^{-\alpha}k^2}\, ,
\ee %
where $D_{\alpha}=R^2/\tau^{\alpha}$ is a generalized diffusion
coefficient. Using the Laplace transform of the fractional
integration
\[\hat{\clL}\left[\prt_t^{-\nu}f(t)\right]= \hat{\clL}
\frac{1}{\Gamma(\nu)} \int_0^t\frac{f(\tau)d\tau}{(t-\tau)^{1-\nu}} =
s^{-\nu}f(s)\,  ,~~~\nu>0\, ,\]
one  obtains the fractional integral equation
\be\label{fieq}
\clP(x,t)-\clP_0(x)=\prt_t^{-\alpha}D_{\alpha}\prt_x^2\clP(x,t)\, .
\ee
Differentiating this equation with respect to time, one obtains
that the CTRW is described by the fractional Fokker-Planck equation
(FFPE) \cite{add2}
\be\label{ffpe} %
\prt_t\clP(x,t)-D_{\alpha}\prt_t^{1-\alpha}\prt_x^2\clP(x,t)=0\, ,
\ee %
where $\prt_t^{\nu}$ is a designation of the Riemann-Liouville fractional
derivative
\[\prt_t^{\nu}f(t)=\frac{d}{dt}\prt_t^{\nu-1}f(t)=
\frac{1}{\Gamma(1-\nu)}\frac{d}{dt}
\int_0^t\frac{f(\tau)d\tau}{(t-\tau)^{\nu}}\,  ,\] %
where $0<\nu<1$ .
From Eq. (\ref{ffpe}) one obtains for the second moment $\lgl
x^2(t)\rgl=\int_{-\infty}^{\infty}x^2\clP(x,t)dx$ the differential
equation:
\[\frac{d}{dt}\lgl x^2(t)\rgl =
\frac{2D_{\alpha}t^{\alpha-1}}{\Gamma(\alpha)}\, .\] %
Here $\Gamma(z)$
is the gamma function, $x(t=0)=0$, and we use the following
property of the fractional derivative
$\prt_t^{\nu}1=t^{-\nu}/\Gamma(1-\nu)$. Therefore, Eq.
(\ref{ffpe}) describes subdiffusion
\be\label{subdif} %
\lgl x^2(t)\rgl=\frac{2D_{\alpha}t^{\alpha}}{\Gamma(1+\alpha)}\,
, \ee %
with the transport exponent $\alpha=0.4$. In the recent numerical
studies of the discrete NLSE \cite{Pikovsky,fks} the exponent
$\alpha$ was found in the range $0.3\div 0.4$.

This consideration can be extended on the wave packet spreading in
the framework of the generalized nonlinear Schr\"odinger equation
\be\label{gnlse}    %
 i\prt_t\psi=\hH\psi+\beta|\psi|^{2n}\psi\, ,
\ee %
where $n\ge 1$ and the Hamiltonian $\hH$ has the pure point
spectrum with the localized eigenfunctions:
$\hH\Psi_k=\clE_k\Psi_k$. For example, the Hamiltonian describes
Wannier-Stark localization \cite{EminHart}, and the discrete
counterpart of Eq. (\ref{gnlse}) with $n=1$ corresponds to
delocalization in a nonlinear Stark ladder \cite{stark1,stark2}.
Repeating  probabilistic consideration of the CTRW based on the
overlapping integrals $A({\bf k})$ of $2n+2$ eigenfunctions
$\Psi_k(x)$, one obtains that Eq. (\ref{gnlse}) describes
subdiffusion with the transport exponent
\be\label{trexp}   %
\alpha=\frac{2}{3+2n}\, .     %
\ee  %
For $n=1$ this expression coincides with the numerical result of
Ref. \cite{stark1}.

The nonlinear Schr\"odinger equation in the presence of disorder
is considered, and the dynamics of an initially localized wave
packet is studied. A subdiffusive spreading of the wave packed is
explained in the framework of a continuous time random walk. It is
shown that the subdiffusive spreading of the initially localized
wave packet is due to the transitions between those AMs which are
strongly overlapped. This overlapping leads to the clustering with
an effective potential well and, correspondingly, to effective
trapping of the wave packet inside each cluster by the potential.
Therefore, the dynamics of the wave packet corresponds to the
CTRW, where the basic mechanism of subdiffusion is an entrapping
of the wave packet with delay, or waiting, times
distributed by the power law $w(t)\sim 1/t^{1+\alpha}\, ,
~~0<\alpha<1$. The trapping mechanism also determines the
transport exponent $\alpha$, which is due to the number of AMs
contributed to the overlapping integrals according to Eq.
(\ref{trexp}). One should recognize that the PDF of the CTRW in Eq.
(\ref{pdf_clP}) is associated with the true quantum distribution
of the NLSE by virtue of Eq. (\ref{trueWF}) with the formal
solution of Eq. (\ref{am10_a}). These expressions are an important
intermediate result which establishes a relation between the NLSE
and the probabilistic description.
Note, that the PDF $\clP(x,t)$ in Eq. (\ref{trueWF})
is the exact distribution, and it corresponds to the CTRW in the long
time and the large scale asymptotics of Eq. (\ref{asympt}).

It is worth noting, in conclusion, that the linear property of
the Liouville equation (\ref{trueWF})  and its formal solution of
Eq. (\ref{am10_a}) are important for the probabilistic approach.
This linear property was used for the CTRW approach, and the
Montroll-Weiss Eq. (\ref{MW}) determines the Green function in Eq.
(\ref{masteq}).

This work was supported in part by the Israel Science Foundation
(ISF), by the US-Israel Binational Science Foundation (BSF).


\begin{thebibliography}{10}
\bibitem{add1} It is worth noting that when $\beta=0$, the Liouville
operator is
$\hL=-\sum_k\omega_k\frac{\prt}{\prt\phi_k}$, where the relation
$a_k=|a_k|e^{i\phi_k}$ is used. Therefore, the Green function is
$\clG=\exp\left(-\sum_k\omega_kt\frac{\prt}{\prt\phi_k}\right)\delta(x-x')$,
and the probability amplitude corresponds to localization
$\clP(x,t)=\sum_{k,l}a^{*}_ka_le^{i(\omega_k-\omega_l)t}\Psi_k(x)\Psi_l(x)$.
\end{thebibliography}
\end{document}